\documentstyle[12pt]{article}

\newcommand{\bm}[1]{\mbox{\boldmath $#1$}}

\begin{document}
\begin{flushright}
OUTP-97-03P
\end{flushright}

\begin{center}

\vspace{.4cm}

{\huge The SU(2) instanton and the adiabatic evolution of two Kramers
  doublets}

\vspace{1.3cm}

{\Large M. T. Johnsson} 

\large{
Department of Physics and Astronomy, 

University of Canterbury, Private Bag 4800, 

Christchurch, New Zealand}

\vspace{.7cm}
{\Large I. J. R. Aitchison} 

\large{
Department of Theoretical Physics, 

1 Keble Road, Oxford OX1 3NP,

England}

\end{center}

\vspace{1cm}

\newpage

\vspace{2cm}

\begin{center}
{\bf \Large{Abstract}}
\end{center}
The adiabatic evolution of two doubly-degenerate (Kramers) levels is
considered. The general five-parameter Hamiltonian describing the system
is obtained and shown to be equivalent to one used in the $\Gamma_8
\otimes(\tau_2\oplus\epsilon)$ Jahn-Teller system. It is shown
explicitly that the resulting SU(2) non-Abelian geometric vector
potential is that of the (SO(5) symmetric) SU(2) instanton. Various
forms of the potentials are discussed.

\newpage

\section{Introduction}
\label{secintro}

\noindent
Adiabatic evolution generates remarkable geometrical structures, as
Berry \cite{berry1984} was the first to emphasise. The evolution of a
single non-degenerate state is associated with a geometric U(1) vector
potential, which is a function of the adiabatically changing parameters
{\bm r}. If this state becomes accidentally degenerate in energy with
another state at some point ${\bm r}^\ast$ in parameter space, the
U(1) potential is that of a magnetic monopole situated at
${\bm r}^\ast$ \cite{berry1984}. When the Hamiltonian is restricted
to be real (rather than Hermitian) the U(1) potential is that of a flux
tube \cite{berry1984,aitchison1988}. Examples of both situations are
known in Jahn-Teller systems: the monopole in the $T\otimes\tau_2$
system \cite{obrien1989}, and the flux tube in the $E\otimes\epsilon$
system \cite{ham1987}.

If the evolving state is itself degenerate throughout the evolution, the
associated vector potential is non-Abelian
\cite{anandan1988b,wilczekET1984}. A natural question to ask then is the
following. Suppose two such doublets become accidentally degenerate
(four-fold degeneracy in all) at some point in parameter space: what
will be the nature of the non-Abelian potentials? The answer to this
question was, in fact, given some time ago \cite{avronET1988}: namely
the potentials are those of the SU(2) Yang-Mills instanton
\cite{belavinET1975,ref9,ref10}. However, the elegant mathematics of
\cite{avronET1988} did not descend to the explicit construction of the
instanton potentials, which are the quantities most physicists like to
deal with. Indeed, since such a degeneracy has co-dimension five (the
geometric Hamiltonian depending on five parameters), the relationship of
the five-dimensional potentials to those of the instanton, which is
normally thought of as living in four-dimensional Euclidean space, is
not completely self-evident. Finally, no specific physical example was
considered in \cite{avronET1988}.

The purpose of the present paper is to fill these gaps. In Section 2 we
briefly recapitulate the case of a two-level crossing and the associated
U(1) monopole potential, in order to bring out later the very close
analogy with the instanton. In Section 3 we obtain the generic
five-parameter Hamiltonian describing this degeneracy pattern, and
observe that it is equivalent to that used in the $\Gamma_8\otimes
(\tau_2\oplus\epsilon)$ Jahn-Teller system. In Section 4 we calculate
the associated geometric vector potentials in a five-dimensional
Cartesian basis, and show --- using the formalism of Jackiw and Rebbi
\cite{jackiwET1976} --- how they are in fact identical to the familiar
four-dimensional instanton potentials. In Section 5 we adopt the
coordinate system used by Yang \cite{yang1978} in his detailed study of
the SU(2) instanton (which he called a generalised monopole), and show
once more that the adiabatically generated potentials agree exactly with
Yang's.

\section{The U(1) monopole and two-level crossing}
\label{secu(1)monopole}
 
\noindent
The generic Hermitian Hamiltonian for any system with accidental
two-level crossing involves three parameters ${\bm r}=(r_1,r_2,r_3)$
and has the form
\begin{equation}
H={\bm r}\cdot{\bm \sigma}
\label{eqHdotsigma}
\end{equation}
where ${\bm \sigma}=(\sigma_1,\sigma_2,\sigma_3)$ are the Pauli
matrices. The eigenvalues of (\ref{eqHdotsigma}) are $\pm r$, where
$r=|{\bm r}|$. One choice of normalised eigenvectors is
\begin{equation}
\psi^+_N=\frac{1}{[2r(r+r_3)]^{\frac{1}{2}}}\left( \begin{array}{c}
r+r_3 \\ r_1+ir_2 \end{array} \right ) = \left( \begin{array}{c}
\cos\frac{\theta}{2} \\ e^{i\phi} \sin\frac{\theta}{2} \end{array}
\right ) 
\label{eqpsi+N}
\end{equation}
corresponding to the eigenvalue $+r$, and
\begin{equation}
\psi^-_N=\frac{1}{[2r(r+r_3)]^{\frac{1}{2}}}\left( \begin{array}{c}
-r_1+ir_2 \\ r+r_3 \end{array} \right ) = \left( \begin{array}{c}
-e^{-i\phi}\sin\frac{\theta}{2} \\ \cos\frac{\theta}{2} \end{array}
\right ) 
\label{eqpsi-N}
\end{equation}
corresponding to the eigenvalue $-r$. In (\ref{eqpsi+N}) and
(\ref{eqpsi-N}) we have given the forms in both Cartesian coordinates
${\bm r}=(r_1,r_2,r_3)$ and in spherical polars
${\bm r}=(r,\theta,\phi)$.

The geometric vector potential $A_a$ is defined by $A_a=\langle\psi|
i\partial_a |\psi\rangle$, where the index $a$ runs over the number of
parameters. In the present case, a short calculation gives
\begin{equation}
{\bm A}^{\pm}_N \equiv\langle\psi^{\pm}_N|i\nabla|\psi^{\pm}_N
\rangle = \frac{\mp 1}{2r(r+r_3)}(-r_2,r_1,0)
\label{eqA_N1}
\end{equation} 
or
\begin{equation}
({\bm A}^{\pm}_N)_{\phi} = \frac{\mp(1-\cos\theta)}{2r\sin\theta}. 
\label{eqA_N2}
\end{equation}
The potentials (\ref{eqA_N1}) and (\ref{eqA_N2}) are those of a magnetic
monopole of strength $\mp \frac{1}{2}$ \cite{ref10,aitchison1987},
located at the level-crossing point $r=0$. The potentials
${\bm A}^{\pm}_N$ are evidently singular at $\theta=\pi$, and the
corresponding eigenvectors are ill-defined at that point. As is well
known \cite{hong-moET1993} this is a consequence of the fact that the
potential for a monopole must be singular on at least one continuous
line running from the monopole to infinity (the Dirac string). To avoid
the singularity one can cover the sphere $S^2$ with two coordinate
patches and define a non-singular vector potential in each patch. The
potentials are linked by a gauge transformation in the region where the
patches overlap. As the notation implies, in the present case the
potentials ${\bm A}_N$ are non-singular over all the surface of $S^2$
except for the south pole $\theta=\pi$. Correspondingly, one can obtain
potentials which are non-singular except at the north pole $\theta=0$ by
using the eigenvectors
\begin{equation}
\psi^+_S=\frac{1}{[2r(r-r_3)]^{\frac{1}{2}}}\left( \begin{array}{c}
r_1-ir_2 \\ r-r_3 \end{array} \right ) = \left( \begin{array}{c}
e^{-i\phi}\cos\frac{\theta}{2} \\ \cos\frac{\theta}{2} \end{array}
\right ) 
\label{eqpsi+S}
\end{equation}
and
\begin{equation}
\psi^-_S=\frac{1}{[2r(r-r_3)]^{\frac{1}{2}}}\left( \begin{array}{c}
-r+r_3 \\ r_1+ir_2 \end{array} \right ) = \left( \begin{array}{c}
-\sin\frac{\theta}{2} \\ e^{i\phi} \cos\frac{\theta}{2} \end{array}
\right ). 
\label{eqpsi-S}
\end{equation} 
The ``S'' potentials are (see also \cite{aitchison1987})
\begin{equation}
{\bm A}^{\pm}_S \equiv\langle\psi^{\pm}_S|i\nabla|\psi^{\pm}_S
\rangle = \frac{\mp 1}{2r(r-r_3)}(r_2,-r_1,0)
\label{eqA_S1}
\end{equation} 
or
\begin{equation}
({\bm A}^{\pm}_N)_{\phi} = \frac{\mp(-1-\cos\theta)}{2r\sin\theta}. 
\label{eqA_S2}
\end{equation}
${\bm A}^{\pm}_N$ are therefore the potentials in the ``northern
hemisphere'' patch, and ${\bm A}^{\pm}_S$ those in the ``southern
hemisphere'' patch. From the spherical polar forms of (\ref{eqpsi+N}),
(\ref{eqpsi-N}), (\ref{eqpsi+S}) and (\ref{eqpsi-S}), we see immediately
that the $\psi^{\pm}_N$ are related to the $\psi^{\pm}_S$ by a phase
transformation 
\begin{equation}
\psi^{\pm}_N=e^{\pm i\phi}\psi^{\pm}_S,
\label{eqNtoS}
\end{equation}
implying that ${\bm A}^{\pm}_N$ and ${\bm A}^{\pm}_S$ are related
by a gauge transformation
\begin{equation}
{\bm A}^{\pm}_N-{\bm A}^{\pm}_S = \mp\nabla\phi,
\end{equation}
which is consistent with (\ref{eqA_N2}) and (\ref{eqA_S2}). If we take
the equator $\theta=\pi/2$ as the overlap region between the $N$ and $S$
patches, we see that after a full circuit of the equator the geometrical
phase $\exp[i\oint{\bm A}\cdot d{\bm r}]$ matches smoothly (via
(\ref{eqNtoS})) from $N$ to $S$, but the non-trivial nature of the gauge
transformation (\ref{eqNtoS}) means that in mathematical language the
U(1) bundle over $S^2$ is non-trivial, and is indeed the monopole
bundle.

\section{The Hamiltonian for the crossing of two doublets}

\noindent
We require a situation in which the two doublets remain degenerate
through adiabatic evolution. This can be ensured only by an appropriate
symmetry, and the natural one to consider here is time-reversal
symmetry. If a system is even under time-reversal and has half-odd
integral total angular momentum, then each energy eigenstate will be at
least doubly degenerate (Kramers degeneracy). We therefore consider a
pair of levels each of which is a Kramers doublet, and construct the
most general Hamiltonian, $H$, describing such a system.

The $4\times 4$ matrix representation of $H$ must be Hermitian, and we
choose a basis such that $H$ is traceless, making the two doublets
degenerate at zero energy. Further, we let $T$ denote the time reversal
operator and $|\phi\rangle$, $|\bar{\phi} \rangle \equiv T|\phi\rangle$,
$|\psi\rangle$, $|\bar{\psi}\rangle$ represent the two Kramers doublets
where $T^2=-1$ and $THT^{-1}=H$.

These equations lead to the relations
\begin{eqnarray}
\langle\phi|H|\bar{\phi}\rangle &=& 0 \\
\langle\phi|H|\phi\rangle &=& \langle\bar{\phi}|H|\bar{\phi}\rangle \\
\langle\phi|H|\psi\rangle &=& \langle\bar{\phi}|H|\bar{\psi}\rangle
^{\ast} \\ 
\langle\phi|H|\bar{\psi}\rangle &=& -\langle\bar{\phi}|H|\psi\rangle
^{\ast}. 
\end{eqnarray}
These constraints lead to a five-parameter description of the
Hamiltonian in the basis $\{ |\phi\rangle, |\bar{\phi}\rangle,
|\psi\rangle, |\bar{\psi}\rangle \}$:
\begin{equation}
H=\left(
\begin{array}{cccc}
r_5 & 0 & r_3+ir_4 & r_1+ir_2 \\
0 & r_5 & -r_1+ir_2 & r_3-ir_4 \\
r_3-ir_4 & -r_1-ir_2 & -r_5 & 0 \\
r_1-ir_2 & r_3+ir_4 & 0 & -r_5 
\end{array}
\right ).
\label{eqH}
\end{equation}
We note that this Hamiltonian can be identified with the Hamiltonian of
\cite{apselET1992} in their consideration of the $\Gamma_8\otimes(
\tau_2\oplus\epsilon)$ Jahn-Teller system. To do this we interchange two
of their basis states:
\begin{equation}
(|1\rangle, |2\rangle, |3\rangle, |4\rangle)\leftrightarrow
(|1\rangle, |4\rangle, |3\rangle, |2\rangle)
\end{equation}
and identify
\begin{eqnarray}
\nonumber
r_1 &=& V_T\cos\beta\sin\theta\cos\phi \\
\nonumber
r_2 &=& -V_T\cos\beta\sin\theta\sin\phi \\
\nonumber
r_3 &=& -V_E\sin\beta\cos\chi \\
\nonumber
r_4 &=& -V_E\sin\beta\sin\chi \\
r_5 &=& V_T\cos\beta\cos\theta.
\end{eqnarray}

Thus we have an interesting physical example in which the non-Abelian
geometrical structure to be discussed in the following sections can be
explored.

\section{The SU(2) instanton and two-doublet crossing}

\noindent
The matrix (\ref{eqH}) has eigenvalues $R,R,-R,-R$ where
$R=(r_1^2+r_2^2+r_3^2+r_4^2+r_5^2)^{\frac{1}{2}}$, so we have the
natural generalization of (\ref{eqHdotsigma}) to the case in which the
levels with energies $+R$ and $-R$ (which cross at $R=0$) are each
doubly degenerate. When the adiabatically evolving level is itself
degenerate, the geometric vector potential becomes a matrix-valued field
(non-Abelian potential) \cite{anandan1988b,wilczekET1984} defined by
\begin{equation}
A^{ij}_a = \langle\psi_j|i\partial_a|\psi_i\rangle
\label{A_ij}
\end{equation}
where $i,j$ run over the labels of the locally single-valued basis in
the degenerate space. We proceed to calculate (\ref{A_ij}) for the
problem defined by (\ref{eqH}). 

One choice of normalised eigenvectors is
\begin{equation}
\psi^+_1=\frac{i}{\sqrt{2R(R-r_5)}} \left [
\begin{array}{c}
r_3+ir_4  \\ -r_1+ir_2  \\ R-r_5 \\ 0 
\end{array} \right ],
\psi^+_2 = \frac{i}{\sqrt{2R(R-r_5)}} \left [
\begin{array}{c}
r_1+ir_2 \\ r_3-ir_4  \\ 0 \\ R-r_5 
\end{array} \right ],
\label{eqcartev+}
\end{equation}
corresponding to the eigenvalue $+R$, and
\begin{equation}
\psi^-_1 = \frac{i}{\sqrt{2R(R+r_5)}} \left [
\begin{array}{c}
-r_3-ir_4  \\ r_1-ir_2  \\ R+r_5 \\ 0 
\end{array} \right ],
\psi^-_2 = \frac{i}{\sqrt{2R(R+r_5)}} \left [
\begin{array}{c}
-r_1-ir_2  \\ -r_3+ir_4  \\ 0 \\ R+r_5 
\end{array} \right ].
\label{eqcartev-}
\end{equation}
corresponding to the eigenvalue $-R$. Inserting (\ref{eqcartev+}) and
(\ref{eqcartev-}) into (\ref{A_ij}) we obtain the potentials
\begin{equation}
A_{a}^{\pm}=\frac{1}{2R(R\mp r_5)} \left[
\begin{array}{c}
r_4\sigma_1 +r_3\sigma_2-r_2\sigma_3 \\ -r_3\sigma_1 + r_4\sigma_2
+r_1\sigma_3  \\
r_2\sigma_1 -r_1\sigma_2+r_4\sigma_3 \\ -r_1\sigma_1 - r_2\sigma_2
-r_3\sigma_3  \\
0 \end{array} \right]
\label{eqsu2cart}
\end{equation}
where the first row on the right hand side of (\ref{eqsu2cart}) gives
the matrix for $A^{\pm}_1$ and so on, ending with $A^{\pm}_5=0$. We note
some similarity with (\ref{eqA_N1}) and (\ref{eqA_S1}). In the following
section we shall see, using a different coordinate system, that the
eigenvectors and geometric potentials are in fact independent of $R$ ---
just as, in the U(1) case, the corresponding quantities in the spherical
basis were independent of $r$. Thus our non-Abelian potentials
(\ref{eqsu2cart}) are naturally defined on the sphere $S^4$. To exploit
this we project from five dimensions onto the surface of the unit
four-dimensional hypersphere via the coordinate transformation
\begin{eqnarray}
r_{\mu} &=& \frac{2x_{\mu}}{1+x^2} \\
r_5 &=& \frac{1-x^2}{1+x^2}
\end{eqnarray}
where $\mu$ runs from 1 to 4 and $x^2=x_{\mu}x_{\mu}$. We then obtain
\begin{equation}
A_{a}^{+}=\frac{1}{2x^2} \left[
\begin{array}{c}
x_4\sigma_1 +x_3\sigma_2-x_2\sigma_3 \\ -x_3\sigma_1 + x_4\sigma_2
+x_1\sigma_3  \\
x_2\sigma_1 -x_1\sigma_2+x_4\sigma_3 \\ -x_1\sigma_1 - x_2\sigma_2
-x_3\sigma_3  \\
0 \end{array} \right]
\label{eqsu2cart2}
\end{equation}
while $A^{-}_{a} = x^2 A_{a}^{+}$, and $A_{5}^{\pm}=0$.

To show that they are indeed SU(2) instanton potentials, we refer to the
paper by Jackiw and Rebbi \cite{jackiwET1976}, which discusses the O(5)
properties of the instanton. They show that the conventional
4-dimensional potentials $\tilde{A}_{\mu}$ are related to our
$A_{\mu}$'s by
\begin{equation}
\tilde{A}_{\mu}=\frac{2}{1+x^2}A_{\mu}
\end{equation}
in the present case (note that our \,$\tilde{ }$\, notation is the
opposite of that in \cite{jackiwET1976}). If we now finally make the
coordinate transformation $x_4\rightarrow -x_4$, we find that our
$\tilde{A}_{\mu}^{-}$ are precisely the negative of the SU(2) instanton
fields defined by \cite{jackiwET1976} and \cite{ref9}:
\begin{equation}
\frac{1}{2}{\bm \sigma}\cdot{\bm A}_{\mu}^{\mbox{inst}} =
\frac{2}{1+x^2}\Sigma_{\mu\nu}x_{\nu}
\end{equation}
where $\Sigma_{\mu\nu}=\eta_{i\mu\nu}\sigma_i/2$ with
$\eta_{i\mu\nu}=-\eta_{i\nu\mu}=\epsilon_{i\mu\nu}$ for $\mu,\nu=1,2,3$
and $\eta_{i\mu\nu}=\delta_{\mu\nu}$ for $\nu=4$. The
$\tilde{A}^{+}_{\mu}$ fields differ from the $\tilde{A}^{-}_{\mu}$
fields by a gauge transformation \cite{ref9}.

The demonstration that the geometric vector potentials in the present
case are just those of the SU(2) instanton is our main result. However,
it is instructive to look at the problem in another way, which casts
further light on the geometry.

We recall that in the monopole case, we needed at least two coordinate
patches to avoid singularities in the vector potential, and that the
potentials were connected at the $S^1$ boundary between the patches by a
non-trivial gauge transformation. Indeed, the associated transition
function \cite{wuET1975,ref10} $\exp[\pm i\phi]$ defines a map from the
$S^1$ equator to the U(1) (structure) group, with winding number $\pm 1$
(and similarly for monopoles of higher magnetic charge). In the
instanton case, $S^4$ can be covered by two patches with an overlap
region which is $S^3$, and the gauge transformation which connects the
two corresponding potentials in this $S^3$ provides a map from $S^3$ to
SU(2) \cite{ref10,yang1978}. These maps are characterised by an integer,
the instanton number. This (topological) number is quite analogous to
the magnetic charge carried by the monopole, but while the latter is
defined via a two-dimensional surface integral of the second-rank field
strength tensor, the former involves a four-dimensional surface integral
of a fourth-rank tensor, namely
$\mbox{Tr}(F_{\mu\nu}F_{\rho\sigma}^{\ast})$, where $F^{\ast}$ is the
$\epsilon$-dual of $F$.

To bring out the interesting role of $S^3$, and of the patches on $S^4$
--- and hence to exploit the U(1) monopole analogy further --- we now
consider our problem using a coordinate system introduced by Yang
\cite{yang1978}. He, incidentally, referred to these configuations as
generalizations of Dirac's monopole. And in the present case, of course,
the configurations are entirely in Euclidean space, and there is no
question of interpreting them as tunnelling events in Minkowski space.

\section{Yang's potentials}

\noindent
Yang uses the following coordinate system:
\begin{eqnarray}
r_i &=& \frac{2R\xi_i\sin\theta}{1+\xi^2} \,\,\,\, i=1,2,3 \\
r_4 &=& \frac{R(1-\xi^2)\sin\theta}{1+\xi^2} \\
r_5 &=& R\cos\theta \\
R &=& (r_i r^i)^{\frac{1}{2}} 
\end{eqnarray}
giving the metric
\begin{equation}
ds^2=dR^2+R^2 d\theta^2+\frac{4R^2\sin^2\theta}{(1+\xi^2)^2}d\xi^2.
\end{equation} 
Note that Yang also uses what he calls ``tensor notation'', where he
ignores the coefficients of the metric, e.g. takes the gradient operator
in spherical polars as $(\partial_r,\partial_{\theta},\partial_{\phi})$
rather than 
$(\partial_r, 1/r \partial_{\theta}, 1/(r\sin\theta)\partial_{\phi})$.

Applying an overall sign change (guided by the previous result) and
setting $X_j=-\frac{i}{2}\sigma_j$ Yang's potentials are
\begin{eqnarray}
A_1^{\alpha} &=& 0   \\   \label{eqAyang_1}
A_2^{\alpha} &=& 0 \\
A_3^{\alpha} &=& \kappa(\frac{1}{2}(1+\xi_1^2-\xi_2^2-\xi_3^2)X_1+
(\xi_1\xi_2-\lambda\xi_3)X_2 + (\xi_1\xi_3+\lambda\xi_2)X_3)   \\
A_4^{\alpha} &=& \kappa((\xi_1\xi_2+
\lambda\xi_3)X_1+\frac{1}{2}(1-\xi_1^2+\xi_2^2-\xi_3^2)X_2+(\xi_2\xi_3-
\lambda\xi_1)X_3)  \\
A_5^{\alpha} &=& \kappa((\xi_1\xi_3-
\lambda\xi_2)X_1+(\xi_2\xi_3+\lambda\xi_1)X_2 +\frac{1}{2}
(1-\xi_1^2-\xi_2^2+\xi_3^2)X_3) 
\label{eqAyang_5}
\end{eqnarray} 
where $\kappa=4i(\mu\cos\theta-\lambda)/(1+\xi^2)^2$.
$\mu=+1,\lambda=+1$ corresponds to Yang's region (or coordinate patch)
$a$ and $\mu=+1,\lambda=-1$ corresponds to region $b$. The region $a$
includes the ``north pole'' $\theta=0$, and the region $b$ includes the
``south pole'' $\theta=\pi$. We shall call these regions $N$ and $S$
respectively. A second, gauge-inequivalent field $A^{\beta}$ is given in
region $N$ by letting $\mu=-1,\lambda=-1$ and in the region $S$ by
letting $\mu=-1,\lambda=+1$. (This is, as Yang shows, the
anti-instanton.)

To obtain these potentials as the geometric vector potentials for our
problem we need to rewrite the Cartesian eigenvectors (\ref{eqcartev+}),
(\ref{eqcartev-}) in terms of Yang's coordinates. Letting
$\gamma^{\mp}=(i\sqrt{2(1\mp\cos\theta)}(1+\xi^2))^{-1}$ the
eigenvectors become
\begin{equation}
\psi_1^{\pm}=\gamma^{\mp} \left [
\begin{array}{c}
\pm \sin\theta(2\xi_3+i(1-\xi^2))  \\
\pm 2\sin\theta(-\xi_1+i\xi_2)  \\
(1+\xi^2)(1\mp\cos\theta)  \\ 0 
\end{array}
\right ]
\end{equation}
\begin{equation}
\psi_2^{\pm}=\gamma^{\mp} \left [
\begin{array}{c}
\pm 2\sin\theta(\xi_1+i\xi_2)  \\
\pm \sin\theta(2\xi_3-i(1-\xi^2))  \\
0 \\ (1+\xi^2)(1\mp\cos\theta) 
\end{array}
\right ],
\end{equation}
corresponding to the eigenvalues $\pm R$ respectively. Now, using
(\ref{A_ij}) with the index $a$ now running over
$\xi_1,\xi_2,\xi_3,\theta$ and $R$, and comparing with Yang's fields
(\ref{eqAyang_1})--(\ref{eqAyang_5}) we obtain
\begin{eqnarray}
A_{a}^+ &=& A_{a}^{(\alpha,S)}
\label{eqmyA1}   \\
A_{a}^- &=& A_{a}^{(\beta,N)}.
\label{eqmyA2}
\end{eqnarray}

The gauge potential $A_{a}^{\alpha}$ so far obtained is defined over
only the $S$ coordinate patch, and the potential $A_{a}^{\beta}$ over
only the $N$ patch. For a full description of the monopole we also need
these potentials in the other patches, namely $A_{a}^{(\alpha,N)}$ and
$A_{a}^{(\beta,S)}$.  Gauge potentials in different patches are related
by a non-Abelian gauge transformation of the form
\begin{equation}
A_{\mu}\rightarrow A'_{\mu}=S(x)A_{\mu}(x)S^{-1}(x)-\frac{i}{g}
(\partial_{\mu} S(x)) S^{-1}(x)
\label{eqgaugetransform}
\end{equation}
where $S$ is an element of the gauge group, in this case SU(2).

In the present case, we may associate a gauge transformation of the
potentials with a unitary transformation $\Lambda$ applied to the basis
vectors in each degenerate subspace:
\begin{eqnarray}
|\psi_i\rangle\rightarrow|\psi'_i\rangle &=& \Lambda_{ij}
 |\psi_j\rangle \\
A^{ij}_{a}\rightarrow A^{ij'}_{a} &=& \langle\psi'_j|i\partial_{a}
|\psi'_i\rangle   \label{eqaprime}  \\
&=& \Lambda A_{a} \Lambda^{-1}+i(\partial_{a}\Lambda)\Lambda^{-1}. 
\end{eqnarray}
Since the intersecting Kramers doublets do indeed describe the SU(2)
instanton, we expect that the other potentials $A_{a}^{(\alpha,N)}$
and $A_{a}^{(\beta,S)}$ should arise from a different choice of basis
vectors.

Both $A_{a}^{(\alpha,S)}$ and $A_{a}^{(\beta,N)}$ are gauge transformed
to their other patch counterparts by (\ref{eqgaugetransform}) with
\cite{yang1978}
\begin{equation}
S=(1-\xi^2+2i{\bm \xi}\cdot{\bm \sigma})/(1+\xi^2).
\end{equation}
Thus we apply the basis change $\Lambda = S$ to the basis vectors
$|\psi_i^{\pm}\rangle$ to obtain an alternative basis set
\begin{equation}
\psi_1^{'\pm}=\gamma^{\mp} \left [
\begin{array}{c}
\mp \sin\theta(1+\xi^2)  \\
0 \\ i(1-\xi^2+2i\xi_3)(1\mp\cos\theta)  \\
-2(1\mp\cos\theta)(\xi_1-i\xi_2)
\end{array}
\right ]
\end{equation}
\begin{equation}
\psi_2^{'\pm}=\gamma^{\mp} \left [
\begin{array}{c}
0 \\ \pm \sin\theta(1+\xi^2)  \\
-2(\xi_1+i\xi_2)(1\mp\cos\theta)  \\
i(1-\xi^2-2i\xi_3)(1\mp\cos\theta) 
\end{array}
\right ],
\end{equation}
using the previous definition of $\gamma^{\mp}$. When put into
(\ref{A_ij}) these new vectors yield
$A^{'+}_{a}=A_{a}^{(\alpha,N)}$ and
$A^{'-}_{a}=A_{a}^{(\beta,S)}$. 

Thus we have identified the two geometric potentials associated with the
higher and lower energy Kramers doublets exactly (up to a gauge
transformation) with Yang's two gauge-inequivalent SU(2) generalised
monopoles:
\begin{eqnarray}
A^+_{a} &=& A_{a}^{\alpha} \\
A^-_{a} &=& A_{a}^{\beta}. 
\end{eqnarray}

Yang shows explicitly that these instanton fields minimise the
four-dimensional Euclidean Yang-Mills action \cite{ref9}. He also
remarks that since he has proved that his fields $\alpha$ and $\beta$
are the only SO(5) symmetrical SU(2) gauge fields (other than the
trivial case), and since the SU(2) instanton is SO(5) symmetrical when
conformally mapped to $S^4$ \cite{jackiwET1976}, the latter must be
identical with one of his fields $\alpha$, $\beta$ (the anti-instanton
corresponding to the other). We have verified this identity of fields
explicitly by calculating the geometric vector potential associated with
the adiabatic evolution of two Kramers doublets.

\section*{Acknowledgements}
\noindent
We would like to thank Professor Geoff Stedman for useful discussions
regarding this paper. IJRA thanks Dr. Mary O'Brien for discussions
concerning Jahn-Teller systems and Mr. Jakov Pfaudler for several patient
explanations of non-Abelian monopoles and instantons. We also thank a
referee for suggesting that we look more closely at the connection with
instantons. MTJ is grateful to the University of Canterbury for the
assistance of a Canterbury Doctoral Scholarship which supported this
work.

\end{document}